\documentclass{Interspeech2024}[hyperref={colorlinks = true,linkcolor = blue}, xcolor=dvipsnames]
\usepackage{enumitem}   
\usepackage{cite}
\usepackage{overpic} 
\usepackage{float}
\usepackage{array}
\usepackage{graphicx}
\usepackage{multirow}
\usepackage{tabu}                     
\usepackage{multicol}                 
\usepackage{makecell}                 
\usepackage{booktabs}                 
\usepackage{caption}
\usepackage[labelfont=normalfont]{subfig}
\usepackage{booktabs}
\usepackage{array, threeparttable}
\usepackage[font=small,labelfont=bf,labelsep=none]{caption}
\interspeechcameraready

\title{MFF-EINV2: Multi-scale Feature Fusion across Spectral-Spatial-Temporal Domains for Sound Event Localization and Detection}

\name[affiliation={}]{Da}{Mu}
\name[affiliation={*}]{Zhicheng}{Zhang}
\name[affiliation={}]{Haobo}{Yue}

\address{
    School of Artificial Intelligence\\
  Beijing University of Posts and Telecommunications, China}
\email{
\thanks{* Corresponding author.}
\{da.mu, zczhang, hby\}@bupt.edu.cn
}


\begin{document}

\maketitle

\begin{abstract}
    Sound Event Localization and Detection (SELD) involves detecting and localizing sound events using multichannel sound recordings. Previously proposed Event-Independent Network V2 (EINV2) has achieved outstanding performance on SELD. However, it still faces challenges in effectively extracting features across spectral, spatial, and temporal domains. This paper proposes a three-stage network structure named Multi-scale Feature Fusion (MFF) module to fully extract multi-scale features across spectral, spatial, and temporal domains. The MFF module utilizes parallel subnetworks architecture to generate multi-scale spectral and spatial features. The TF-Convolution Module is employed to provide multi-scale temporal features. We incorporated MFF into EINV2 and term the proposed method as MFF-EINV2. Experimental results in 2022 and 2023 DCASE challenge task3 datasets show the effectiveness of our MFF-EINV2, which achieves state-of-the-art (SOTA) performance compared to published methods.
\end{abstract}

\section{Introduction}
Sound Event Localization and Detection (SELD) aims to use multichannel sound recordings to detect the onset and offset of sound events within specific target classes and estimate their direction of arrival (DoA). Its applications cover various fields, including smart homes, surveillance systems, and human-computer interaction. Since the introduction of SELD in the Detection and Classification of Acoustic Scenes and Events (DCASE) challenge as task3\cite{adavanne2018sound}, deep neural network (DNN) based methods have been widely studied. These methods can be implemented using either single-branch or dual-branch neural networks. Single-branch approaches \cite{adavanne2018sound, shimada2021accdoa, shimada2022multi, niu2023experimental, shul2023cst} commonly utilize a combination of log-mel spectrograms and intensity vectors (IVs) as input and output the results in a class-wise format. Dual-branch approaches\cite{Cao2020_task3_report, cao2021improved, Hu2022, hu2022track} divide the network into sound event detection (SED) branch and DoA branch, treating SELD as a multi-task learning problem. The SED branch takes log-mel spectrograms as input, while the DoA branch uses both log-mel spectrograms and IVs as input. These methods usually generate results in a track-wise format \cite{Cao2020_task3_report}. 


The Event-Independent Network V2 (EINV2) \cite{Hu2022} is a dual-branch network architecture that has demonstrated promising performance in the 2022 DCASE challenge. It utilizes four dual convolution (Dual Conv) layers as the encoder and Conformer blocks \cite{gulati20_interspeech} as the decoder. The two branches are connected through soft parameter sharing \cite{misra2016cross}. However, most widely used architectures \cite{Hu2022, hu2022track,   niu2023experimental, shul2023cst} simply stack CNN or ResNet in their encoders, which limits their ability to effectively exploit the spectral, spatial, and temporal domains when extracting features. Research conducted in other fields, such as speech enhancement \cite{hao2021fullsubnet,chen2022fullsubnet+, yang2023mcnet, tolooshams2020channel} and sound event detection \cite{hu2022multi, Nam2023}, has proved the importance of fully utilizing these three domains.

Similarly, we argue that leveraging the spectral, spatial, and temporal domains is crucial for SELD. \textbf{1) Spectral Domain:} The spectral domain offers distinctive cues related to the frequency content of sound events. Each sound event exhibits unique spectral characteristics, encompassing frequency details and broader spectral trends. Leveraging multi-scale spectral information allows the model to focus on both local and global spectral patterns, thereby enhancing the identification of sound events. \textbf{2) Spatial Domain:} The spatial domain provides critical locational information about sound sources. Multichannel audio has the potential to model numerous aspects of spatial scenes, making the simultaneous determination of multiple sound sources possible. Utilizing multi-scale spatial information enables the model to localize sound sources from different directions, which greatly benefits DoA estimation. \textbf{3) Temporal Domain:} The temporal domain provides details about the evolution of sound events over time. This enables the model with the ability to track temporal context dynamics. Exploiting multi-scale temporal information, the model is able to capture the characteristics of target sound events with short- or long-duration, assisting in comprehension of the entire audio segment.

\begin{figure*}[t]
	\centering
	\subfloat[\textnormal{}]{\includegraphics[width=.29\linewidth]{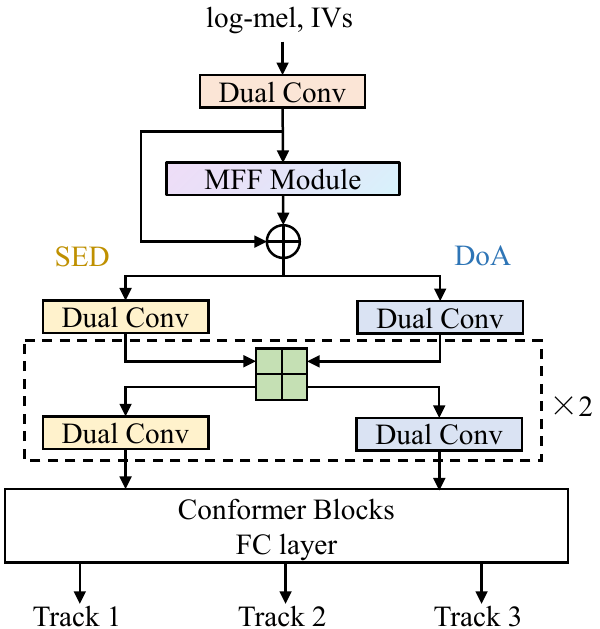} \label{fig1a}}
        \hspace{10pt}
	\subfloat[\textnormal{} \label{fig1b}]{\includegraphics[width=.68\linewidth]{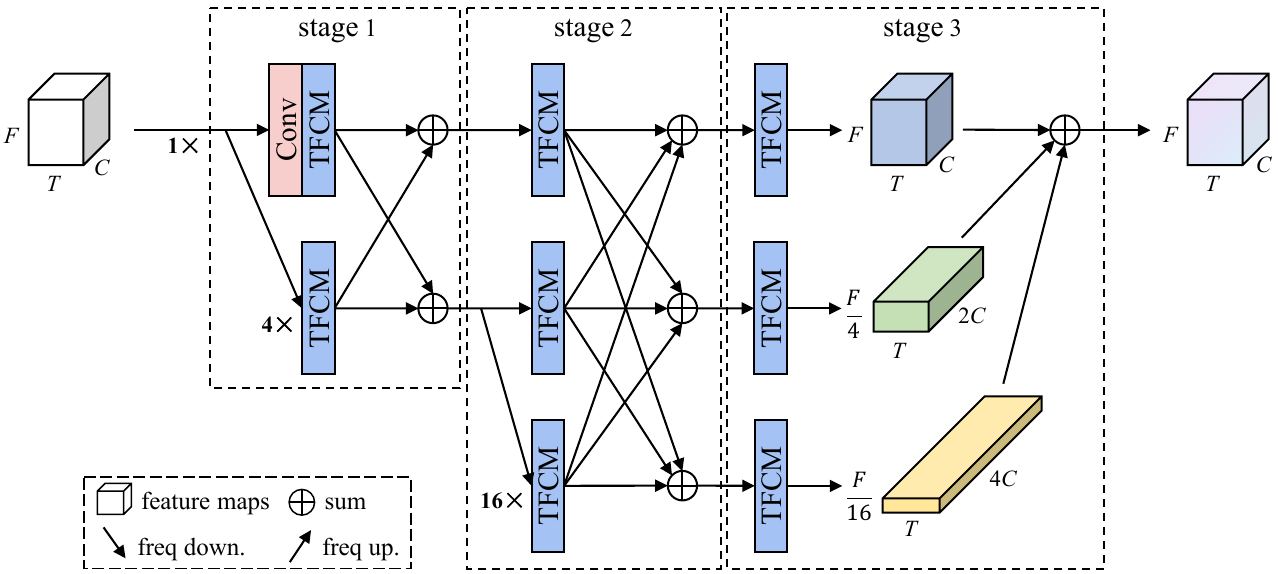}}
	\caption{\textnormal{An illustration of (a) the MFF-EINV2 pipeline diagram and (b) the details of the MFF module. (a) The architecture of the Conformer blocks and FC layer aligns with the EINV2 and the green boxes indicate soft parameter sharing. (b) \textit{C}, \textit{T}, and \textit{F} are the dimension sizes of channel, time, and frequency, respectively.   \textquotedblleft1×", \textquotedblleft4×", and \textquotedblleft16×" represent different scales of feature maps. \textquotedblleft freq down." and \textquotedblleft freq up." refer to frequency downsampling (FD) and frequency upsampling (FU), respectively.}}
        \label{MFF-EINV2 and MFF}
\end{figure*}



In this paper, instead of simply stacking CNN or ResNet in the encoder, we propose a three-stage neural network structure called Multi-scale Feature Fusion (MFF) module to improve the ability to learn multi-scale features across spectral, spatial, and temporal domains. We incorporated the MFF module into EINV2 and term the proposed method as MFF-EINV2. The main contributions of this work are as follows:

\begin{itemize}[
    itemindent = 0pt,
    labelindent = \parindent,
    labelwidth = 2em,
    labelsep = 5pt,
    leftmargin = *]
    \item We employ parallel subnetworks architecture with the TF-Convolution Module (TFCM) \cite{zhang2022multi} to extract multi-scale features across spectral, spatial, and temporal domains. Repeated multi-scale fusion is used to improve the representation capabilities of the subnetworks, enhancing feature extraction.
    \item Our MFF-EINV2 outperforms EINV2 by reducing parameters by 68.5\% and improving the SELD$_{score}$ by 18.2\%. It also surpasses other published methods and achieves state-of-the-art (SOTA) performance.
\end{itemize}

\section{Proposed Method}

The proposed MFF-EINV2 architecture is illustrated in Fig.\ref{MFF-EINV2 and MFF}.\subref{fig1a}. We concatenate log-mel spectrograms and IVs along the channel dimension as input. The first step in the process involves a Dual Conv layer, which transforms the input into a feature map with dimensions [$C$, $T$, $F$], where $C$, $T$, and $F$ represent the dimension size of the channel, time, and frequency, respectively. Next, we introduce the MFF module to fully extract multi-scale features from spectral, spatial, and temporal domains, as shown in Fig.\ref{MFF-EINV2 and MFF}.\subref{fig1b}. The network is then divided into two branches: SED branch and DoA branch. Each branch consists of three Dual Conv layers. After each Dual Conv layer, a pooling layer with a kernel size of (2,2) is applied, which aims to align the temporal resolution with the target label. To prevent the MFF module from losing original information or extracting irrelevant features, a residual connection is established between the output of the MFF module and the first Dual Conv. Our encoder architecture maintains a balance between network complexity and soft connections.

The subsequent network structure is the same as EINV2. The decoder utilizes Conformer blocks, which integrate convolution layers and multi-head self-attention (MHSA) mechanisms to extract both local and global time context information of feature sequence simultaneously. Finally, the fully connected (FC) layer outputs three tracks, enabling the handling of up to three overlapping sound events. In the following sections, we will present the details of our MFF module one by one.

\subsection{Parallel Multi-resolution Subnetworks}
We introduce parallel multi-resolution subnetworks to generate multi-scale spectral and spatial features. The input to the MFF module is a feature map of dimensions [$C$, $T$, $F$]. We utilize frequency downsampling (FD) to generate parallel subnetworks in stage 1 and stage 2. FD is implemented through a block that consists of a 2D convolution (Conv2D) layer, a batch normalization (BN) layer, and a Rectified Linear Unit (ReLU) activation layer. Parameters for Conv2D include a kernel size of (1, 7), stride of (1, 4), and padding of (0, 2), resulting in reducing the frequency dimension by 4 times and expanding the channel dimension by 2 times. Throughout the generation of these subnetworks, we ensure that the time dimension ($T$) remains at a high resolution, which is able to provide more detailed temporal context information. As a result, the feature map dimensions of the second subnetwork become [2$C$, $T$, $F/4$], and the third subnetwork has dimensions [4$C$, $T$, $F/16$], as shown in Fig.\ref{MFF-EINV2 and MFF}.\subref{fig1b}.

\begin{figure}[htbp]
	\centering
	\subfloat[]{\includegraphics[width=.4\linewidth]{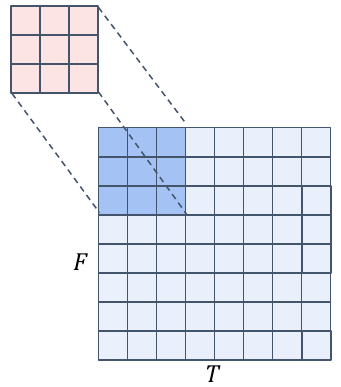} \label{fig2a}}
        \hspace{20pt}
	\subfloat[]{\includegraphics[width=.4\linewidth]{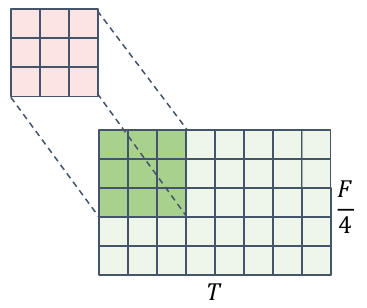} \label{fig2b} }
	\caption{\textnormal{An illustration of the convolution operation on (a) a high-resolution frequency feature map from the first subnetwork and (b) a low-resolution frequency feature map from the second subnetwork. The pink boxes indicate the convolution kernel of D-Conv in TFCM. A green time-frequency (T-F) bin contains the information of seven consecutive blue T-F bins.}} 
    \label{local and global}
\end{figure}

The spectral information of sound events is mainly reflected in the frequency dimension\cite{nam22_interspeech}. The MFF module consists of three subnetworks with different frequency resolutions, each focusing on different scales of spectral patterns. In the first subnetwork, when the feature map passes through the convolution layer in TFCM, the convolution kernel interacts with a specific frequency and its neighboring frequencies, as depicted in Fig.\ref{local and global}\subref{fig2a}. This process allows the model to focus on local spectral patterns and frequency details of sound events. Next, the second subnetwork handles a feature map of dimensions [2$C$, $T$, $F/4$]. Due to the FD operation, each frequency unit now incorporates information from seven original contiguous frequencies. It effectively compresses high-resolution frequency information to a lower resolution. Even though the size of the convolution kernel remains the same, its coverage on the original feature map increases as resolution decreases. This means that the convolution layer is able to model global spectral patterns and broader frequency trends of sound events, as shown in Fig.\ref{local and global}\subref{fig2b}. Similarly, the third subnetwork focuses on more global spectral patterns. This method enables the model to attend to spectral patterns at various scales, leading to improved discrimination and identification of different sound events.

The spatial information of sound sources is encoded in both the channel and frequency dimensions\cite{tolooshams2020channel,wang23j_interspeech,huang2023swg}. FD changes these two dimensions, establishing a complex mapping between feature maps and spatial information. Each subnetwork extracts features from distinct aspects of the spatial domain, promoting a comprehensive understanding of the acoustic scene. This approach improves the model's capacity to simultaneously estimate the DoA of sound sources from different directions.

\subsection{TF-Convolution Module}

We use TFCM in each subnetwork in order to extract multi-scale temporal features. The TFCM is constructed by sequentially concatenating $m$ convolutional blocks. The convolutional block is composed of two pointwise convolution (P-Conv) layers and a 2D depthwise convolution (D-Conv) layer. The D-Conv layer utilizes small convolution kernels of size (3, 3) to maintain computational efficiency. The dilation rate of the D-Conv layer's time dimension increases from 1 to $2^{m-1}$. 

When the dilation rate is low, the TFCM has the ability to gather brief and fine-grained temporal context information, which is especially useful for identifying short-duration sound events. As the dilation rate increases, the TFCM is able to capture extended and coarse-grained temporal context information, which is effective in recognizing long-duration sound events. Additionally, the time evolution of the log-mel spectrograms amplitude potentially indicates the stationarity of sound events\cite{nam22_interspeech}. Therefore, TFCM facilitates multi-scale modeling of signal stationarity, enabling the distinction between stationary and non-stationary signals. This method enables the model to learn the multi-grained temporal context dynamics of sound events, thereby improving the capabilities of continuous SED and DoA estimation.

\subsection{Repeated Multi-scale Fusion}
We employ repeated multi-scale fusion to enhance the representation capacities of each subnetwork by continuously exchanging information with other parallel subnetworks. During stages 1 and 2, each subnetwork merges features from all other subnetworks, enabling a comprehensive fusion. Since analyzing the frequency dimension at high resolution provides more detailed spectral and spatial information, the last two subnetworks are selectively fused with the first subnetwork in stage 3 to restore the frequency dimension to $F$. We use FD and frequency upsampling (FU) to meet the dimension requirements of feature fusion. FD comprises FD×4 and FD×16. FD×16 includes an additional Conv2D layer compared to FD×4, enabling frequency downsampling by a factor of 16 and channel upsampling by a factor of 4. FU consists of FU×4 and FU×16. FU×$n$ changes the channel dimension size using a Conv2D layer with a kernel size of (1,1), followed by applying the nearest neighbor algorithm for $n$ times frequency upsampling, where $n$ is 4 or 16.
For instance, the multi-scale fusion process in stage 3 is as follows:
\begin{equation}
{\textbf{Y$_{1}$}  = \textbf{X$_{1}$} + {\rm FU\times 4}(\textbf{X$_{2}$}) + {\rm FU\times16}(\textbf{X$_{3}$}).}
\end{equation}

\noindent where \textbf{X$_{1}$}$\in\mathbb{R}^{C\times T\times F}$, \textbf{X$_{2}$}$\in\mathbb{R}^{2C\times T \times {F/4}}$, \textbf{X$_{3}$}$\in\mathbb{R}^{4C\times T \times {F/16}}$ represent the feature map of the first, second, and third subnetwork, respectively. \textbf{Y$_{1}$}$\in\mathbb{R}^{C\times T \times {F}}$ denotes the feature map after multi-scale fusion.

Each parallel subnetwork is responsible for processing feature maps at different scales, enabling them to capture various aspects of features in spectral, spatial, and temporal domains. As a result, these subnetworks contain complementary information to each other. Through multi-scale fusion over and over again, these subnetworks facilitate information exchanges, allowing each to simultaneously consider various scale information across the three domains. This repeated fusion process enhances their ability to extract more meaningful features from the input.

\section{Experiments}

\subsection{Datasets}
We trained and evaluated MFF-EINV2 on the STARSS22\cite{Politis2022} and STARSS23\cite{Shimada2023starss23_arxiv} datasets, which are used for 2022 and 2023 DCASE challenge task3, respectively. The difference between the two datasets is that STARSS23 has an additional 2 hours and 30 minutes of recordings in the development set. Both datasets contain thirteen sound event classes and two types of multichannel array signals, including first-order Ambisonics (FOA) and tetrahedral microphone array signals. We trained our model only on FOA array signals, which comprise four channels: an omnidirectional channel($w$), and three directional channels ($x$, $y$, and $z$). In addition, we used \textit{synthetic data}\cite{politis_2022_6406873} as external data. The \textit{synthetic data} was generated by convolving single-channel data from FSD50K\cite{fonseca2021fsd50k} and TAU-SRIRs\cite{politis_2022_6408611}. For both STARSS22 and STARSS23, we utilized the \textit{dev-set-train} and the \textit{synthetic data} for training, and the \textit{dev-set-test} for evaluation. 

\begin{table*}[th]
  \belowrulesep=0pt
  \aboverulesep=0pt
  \caption{\textnormal{Performance comparison of the proposed MFF-EINV2 with the other SELD models on the \textit{dev-set-test} of STARSS22 and STARSS23 datasets.}}
  \centering
  \begin{tabular}{c|c|c|cccc|c}
    \toprule
    Datasets & Models & Params &ER$_{20^\circ}$↓ & F$_{20^\circ}$↑ & LE$_{CD}$↓ & LR$_{CD}$↑ & SELD$_{score}$↓ \\
    \midrule
    \multirow{5}{*}{STARSS22} 
    & 2022 Baseline\cite{Politis2022}   & 0.6M & 0.72 & 24.0 & 26.6 & 49.0 &   0.5345 \\
    & ResNet-Conformer\cite{niu2023experimental} & 13.1M & 0.71 & 31.0 & 22.3 &  {64.0} &   0.4710 \\
    
    & CST-former\cite{shul2023cst}    & - & \textbf{0.59} & \textbf{42.6} & {20.5} &{61.3} &   {0.4162}  \\
    \cline{2-8}
    & EINV2\cite{Hu2022}   & 85.3M & 0.75 & 32.3 & 24.0 & 56.1 &   0.5000  \\
    & MFF-EINV2(proposed) & 26.9M & {0.61} & {41.7} & \textbf{20.3} & \textbf{66.9} &   \textbf{0.4089} \\
    \cline{1-8}
    \multirow{4}{*}{STARSS23} 
    & 2023 Baseline\cite{Shimada2023starss23_arxiv} & 0.7M & 0.57 & 29.9 & 22.0 & 47.7 &   0.4791 \\
    & DST attention\cite{YShul_KAIST_task3a_report} & - & 0.58 & 39.5 & 20.0 & 55.8 &   0.4345 \\
    & CST-former\cite{shul2023cst} & - & {0.56} & \textbf{42.7} & \textbf{17.9} & {62.0} &   {0.4019} \\
    & MFF-EINV2(proposed)  & 26.9M & \textbf{0.54} & {42.5} & {18.7} & \textbf{62.6} & \textbf{0.3980}  \\
    \bottomrule
  \end{tabular}
  \label{STARSS}
\end{table*}

\subsection{Hyper-parameters and Evaluation Metrics}
We processed audio clips into fixed 5-second segments without overlap for both training and evaluation. The sampling rate of the audio is \SI{24}{\kilo\hertz}. We applied a Short-Time Fourier Transform (STFT) with a hop size of 300, using a 1024-point Hanning window. Next, we generated log-mel spectrograms and IVs in the log-mel space, with 128 frequency bins. IVs were concatenated with log-mel spectrograms along the channel dimension, resulting in 7-channel input. After each Dual Conv layer in the encoder, the channel dimension of the feature map is 64, 128, 256, and 256, respectively. The structure of the Conformer blocks remains consistent with EINV2, but the embedding dimension is reduced from 512 to 256. The number of Conformer layers is 2. Loss weight is 0.8 for SED and 0.2 for DoA. We utilized the AdamW optimizer, training for a total of 100 epochs. For both STARSS22 and STARSS23, the initial learning rate is 0.0003. With STARSS22, it drops to 0.00003 after 80 epochs, and for STARSS23, after 60 epochs. We set the batch size to 6 and trained for 43 hours using an NVIDIA GeForce RTX 3090 GPU.

We utilized five metrics for evaluation\cite{adavanne2018sound}. Error rate (ER$_{20^\circ}$) and macro-averaged F1-score (F$_{20^\circ}$) are location-dependent used for SED. Localization error (LE$_{CD}$) and localization recall (LR$_{CD}$) are class-dependent employed for DoA estimation. Furthermore, we calculate the average of these four metrics to obtain SELD$_{score}$.

\subsection{Experimental Results}

\subsubsection{Comparison with other methods}
To investigate the effectiveness of our proposed MFF-EINV2, we compare it with several other models\cite{Politis2022, Hu2022, niu2023experimental,shul2023cst,Shimada2023starss23_arxiv,YShul_KAIST_task3a_report}. 
The 2022 baseline\cite{Politis2022} uses a convolutional recurrent neural network (CRNN) and was updated in 2023\cite{Shimada2023starss23_arxiv} to include two additional MHSA layers. The ResNet-Conformer\cite{niu2023experimental} utilizes ResNet as the encoder and Conformer as the decoder. The DST attention\cite{YShul_KAIST_task3a_report} introduces a spectral attention layer into the decoder, enhancing the performance over the 2023 baseline. The CST-former\cite{shul2023cst} incorporates distinct attention mechanisms to process channel, spectral, and temporal information independently. For our MFF-EINV2, the MFF module contains three stages, while TFCM consists of six convolutional blocks. All models were trained without any data augmentation techniques. The performance comparison is shown in Table \ref{STARSS}.

First, we compare the results on the STARSS22 dataset. Our model significantly outperforms all evaluation metrics of the 2022 Baseline and ResNet-Conformer. Moreover, our model performs slightly better than CST-former, showing a 1.8\% (0.0073) improvement in SELD$_{score}$. It particularly excels in LE$_{CD}$ and LR$_{CD}$, but falls short in ER$_{20^\circ}$ and F$_{20^\circ}$. This may be attributed to more complex and significant features extracted from the spatial domain compared to the spectral domain during the generation of subnetworks. In comparison to EINV2, our model reduces the number of parameters by 68.5\% (58.4M), primarily due to the embedding dimension of the Conformer is reduced from 512 to 256. Nevertheless, our model has an 18.2\% (0.0911) decrease in SELD$_{score}$. This is because our model fully leverages the spectral, spatial, and temporal domains to extract more meaningful features. 

We now compare the performance achieved on the STARSS23 dataset. Our model outperforms 2023 baseline and DST attention by a significant margin in all metrics. Although the lead is modest compared to CST-former, our model achieves the best SELD$_{score}$ and surpasses CST-former in ER$_{20^\circ}$ and LR$_{CD}$. This may be due to our method enhancing the model's focus on both fine details and the overall structure of sound events, improving ER$_{20^\circ}$. Moreover, it integrates spatial information across multiple scales, enhancing the model's understanding of the sound scene and likely leading to a higher LR$_{CD}$.


\begin{table}[th]
  \belowrulesep=0pt
  \aboverulesep=0pt
  \caption{\textnormal{Ablation experiments with the number of parallel subnetworks $s$ on the STARSS22 dataset.}}
  \centering
  \begin{tabular}{c|cccc|c}
    \toprule
    $s$ & ER$_{20^\circ}$↓ & F$_{20^\circ}$↑ & LE$_{CD}$↓ & LR$_{CD}$↑ & SELD$_{score}$↓ \\
    \midrule
    0 & 0.65 &37.6  & 22.6 & 60.5 & 0.4484  \\
    1 & 0.63 & 40.5 & 22.3 & 59.8 & 0.4384  \\
    2 & 0.64 & 38.2 & 19.7 & 61.3 & 0.4379  \\
    3 & \textbf{0.61} & \textbf{41.7} & 20.3 & \textbf{66.9} &   \textbf{0.4089} \\
    4 & 0.68 & 34.1 & \textbf{19.4} & 57.9 & 0.4667  \\
    \bottomrule
  \end{tabular}
  \label{parallel numbers}
\end{table}

\subsubsection{Number of parallel subnetworks}

The number of parallel subnetworks, denoted as $s$, directly impacts the model's representation capability and computational cost. A larger $s$ yields more spectral and spatial feature maps at various scales, thus enhancing the expressiveness of the trained SELD model. However, too large $s$ leads to higher computational costs and might cause overfitting of the model. Table \ref{parallel numbers} shows the SELD performance of MFF-EINV2 with different numbers of parallel subnetworks. From the table, using three parallel networks strikes a balance between representation ability and fitting ability, resulting in the best SELD performance.


\subsubsection{Number of convolutional blocks in TFCM}

The number of convolutional blocks $m$ in TFCM influences the extraction of multi-scale temporal information. A larger $m$ will extract more scales of temporal information. However, too large $m$ might extract irrelevant features due to extensive dilation in D-Conv. Table \ref{TFCM numbers} shows the SELD performance of MFF-EINV2 with different numbers of convolutional blocks in TFCM. From the table, using six convolutional blocks effectively captures multi-scale temporal information.

\begin{table}[th]
  \belowrulesep=0pt
  \aboverulesep=0pt
  \caption{\textnormal{Ablation experiments with the number of convolutional blocks $m$ in TFCM on the STARSS22 dataset.}}
  \centering
  \begin{tabular}{c|cccc|c}
    \toprule
    $m$ & ER$_{20^\circ}$↓ & F$_{20^\circ}$↑ & LE$_{CD}$↓ & LR$_{CD}$↑ & SELD$_{score}$↓ \\
    \midrule
    3 & 0.64 & 38.1 & 21.7 & 61.0 & 0.4414  \\
    4 & 0.66 & 34.8 & 20.3 & 59.3 & 0.4566  \\
    5 &0.66  & 38.3 & 20.5 &64.2  &  0.4375 \\
    6 & \textbf{0.61} & \textbf{41.7} & \textbf{20.3} & \textbf{66.9} &   \textbf{0.4089} \\    
    7 & 0.67 & 37.0 & 20.7 & 65.0 &  0.4414 \\
    \bottomrule
  \end{tabular}
  \label{TFCM numbers}
\end{table}

\section{Conclusion}
In this paper, we propose MFF-EINV2, a novel approach for SELD. In the MFF module, we introduce parallel subnetworks and employ TFCM to extract multi-scale features across spectral, spatial, and temporal domains. In addition, we leverage repeated multi-scale fusion between parallel subnetworks so that each subnetwork continuously receives information from other parallel representations. Results show that compared to EINV2, our MFF-EINV2 significantly reduces parameters by 68.5\% while improving the SELD$_{score}$ by 18.2\%, demonstrating the effectiveness of our MFF module. Notably, without data augmentation, our proposed MFF-EINV2 outperforms other published methods and achieves SOTA performance. 


%

\bibliographystyle{IEEEtran}
\bibliography{ourbib}

\end{document}